\documentclass[twocolumn,showpacs,preprintnumbers,amsmath,amssymb]{revtex4}
\usepackage{epsfig}
\preprint{LBSCO/2006}
\begin{document}
\title{Effects of Superconductivity and Charge Order on the sub-Terahertz reflectivity of La$_{1.875}$Ba$_{0.125-y }$Sr$_{y}$CuO$_4$}
\author{M. Ortolani$^1$, P. Calvani$^1$, S. Lupi$^1$, U. Schade$^2$, A. Perla,$^1$ M. Fujita$^3$ and K. Yamada$^3$ }
\affiliation{$^1$"Coherentia" - INFM and Dipartimento di Fisica, Universit\`a di Roma La Sapienza, Piazzale Aldo Moro 2, I-00185 Roma, Italy}
\affiliation {$^2$Berliner Elektronenspeicherring Gesellschaft f\"{u}r Synchrotronstrahlung m.b.H, Albert-Einstein-Str.15, D-12489 Berlin, Germany}
\affiliation {$^3$Institute for Chemical Research, Kyoto University, Gokasho, Uji 610-0011, Japan}
\date{\today}

\begin{abstract}
The reflectivity $R(\omega)$ of both the $ab$ plane and the $c$ axis of two single crystals of La$_{1.875}$Ba$_{0.125-y }$Sr$_{y}$CuO$_4$ has been measured down to 5 cm$^{-1}$, using coherent synchrotron radiation below 30 cm$^{-1}$. For $y$ = 0.085, a Josephson Plasma Resonance is detected at $T \ll T_c$ = 31 K in the out-of-plane $R_{c}(\omega)$, and a far-infrared peak (FIP) appears in the in-plane optical conductivity $\sigma_{ab} (\omega)$ below 50 K, where non-static CO is reported by X-ray scattering. For $y$ = 0.05 ($T_c$ = 10 K), below the ordering temperature $T_{CO}$, a FIP is again observed in $\sigma_{ab} (\omega)$. The FIP then appears to be an infrared signature of CO, either static or fluctuating, as reported in previous works on the La-Sr cuprates.

\end{abstract}
\pacs{74.25.Gz, 74.72.-h, 74.25.Kc}
\maketitle

\section{Introduction}

The study of cuprates at 1/8 doping attracted considerable interest since when, in the $T_c$ vs. $x$ plot of the first high-$T_c$ superconductor to be discovered, (La$_{2-x}$Ba$_{x}$CuO$_4$ , LBCO), a drop from 30 K to almost zero was found at $x \simeq 0.125$ \cite{Tokura}. The search was extended to other cuprate systems with hole doping close to 1/8 \cite{Tokura2}, but up to now such a strong effect on $T_c$ has been confirmed only in the 214-systems, which include La$_{2-x}$Sr$_{x}$CuO$_4$ (LSCO), LBCO, (La,Nd)$_{2-x}$Sr$_{x}$CuO$_4$ (LNSCO), and La$_{1.875}$Ba$_{0.125-y}$Sr$_y$CuO$_4$ (LBSCO). They undergo a transition at a $T_{d1}$ above room temperature, from the High-Temperature Tetragonal (HTT) phase, where the Cu-O planes have a square symmetry and the average tilting angle $\theta$ of the oxygen octahedra is zero, to the Low-Temperature Orthorombic (LTO) phase, with an average $\theta \simeq 2^o$. In the 214-systems, for fixed hole-doping $x=0.125$, $T_c$ is strongly dependent from the cation composition and ranges from 1 to 33 K \cite{McAllister1}. The key parameter for understanding the $T_c$-drop in the 214-systems seems to be the disorder due to the inhomogeneous cation-size distribution in the cation layer. This allows a tilting of the Cu-O octahedra up to $\theta \sim 4^o$, larger than in the LTO phase, \cite{McAllister2} which triggers a structural phase transition at a low $T_{d2}$ temperature. One finds different $0 \alt T_{d2} \alt 100 K$ for different 214-systems, according to the cation-size distribution. 

At $T_{d2}$ an electronic transition also occurs, which brings the system into a static charge-order (CO) state, with very low $T_c$. This effect was first observed in the Cu-O planes of (La,Nd)$_{2-x}$Sr$_{x}$CuO$_4$ (LNSCO) \cite{Tranquada} and, later, of La$_{1.875}$Ba$_{0.125-y}$Sr$_y$CuO$_4$ (LBSCO).\cite{FujitaPrl} The CO state, from diffraction data, is in form of one-dimensional charge stripes. They act as charged walls which separate antiferromagnetic domains. The above scenario points toward an intrinsic competition between static, long-range CO, and SC states. However, magnetic neutron scattering revealed a spin modulation with doping-dependent wavevector for any $x$ in LSCO,\cite{Yamada,Yamada2} much similar to that of LNSCO \cite{TranquadaPrb} and LBSCO \cite{FujitaPrl}. These and other results \cite{stripes} suggest that i) the magnetic phase separation related to the stripe state survives also when $T_c$ is not depressed; ii) that CO instabilities may be a general feature of the Cu-O plane, independently of the out-of-plane structure. In this scenario, superconductivity may be suppressed in LNSCO and LBSCO by other factors peculiar to these compounds, like disorder, structural transitions or static spin order (SO) \cite{TranquadaXrays}, rather than by the charge order itself. This latter, in form of short-lived CO fluctuations, might instead provide singular interactions between electrons, ultimately leading to SC pairing.\cite{stripes} 

\begin{figure}[b]
{\hbox{\psfig{figure=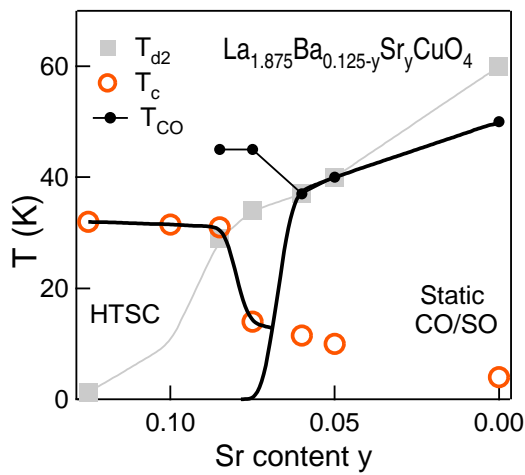,width=8cm}}}
\caption{(Color online) Phase diagram of LBSCO as a function of the Sr content, as elaborated from Ref. \onlinecite{FujitaNew1} The transition temperatures towards the superconducting HTSC phase ($T_c$), the charge/spin ordered state ($T_{CO}$) and the LTT/LTLO structure ($T_{d2}$) are shown. The lines are just guides to the eye.}
\label{diag}
\end{figure}

La$_{1.875}$Ba$_{0.125-y}$Sr$_{y}$CuO$_4$  is the ideal system to observe the effect of the transition at $T_{d2}$ on the electronic properties, and eventually detect fluctuating CO in the vicinity of the high-$T_c$ state. Indeed in LBSCO, unlike in LNSCO, the in-plane structural distortion can be finely tuned by an increasing Ba $\to$ Sr substitution. Therefore $T_{d2}$ could be measured, by either X-ray and neutron diffraction, as a function of $y$ \cite{FujitaPrl,FujitaPrb,FujitaXray,FujitaNew1, FujitaNew2}. While 1/8-doped LSCO remains in the LTO phase at low T, all LBSCO samples with $y < 0.09$ show a low-$T$ structural transition with a $T_{d2}$ ranging from 30 K at $y = 0.09$ to 60 K at $y=0$, as reported in Fig.\ \ref{diag}. A detailed analysis of the diffraction patterns showed that the low-T phase is LTT for $y < 0.075$, the so-called Low-Temperature Less-Ortorhombic (LTLO) phase for $y \geq 0.075$. Calculations based on the Landau-Ginzburg theory \cite{LG} indicate that the LTO-LTT transition is of first order, the LTO-LTLO transition of second order. Both CO and SO superlattice spots have been observed for any $y < 0.09$ below a $T_{CO}$, also shown in Fig.\ \ref{diag}, close to $T_{d2}$ but not identical. In particular, the CO diffraction spots are still observed above $T_{d2}$ in samples with $0.075 < y < 0.09$, to suddenly die out at $T > T_{d2}$ for $y < 0.075$. In conclusion, the LBSCO series shows a clear crossover at $y \sim 0.075$, both in the CO and in the structural patterns. Interestingly, at $y \sim 0.075$ one also observes a drop in $T_c$ vs. $y$ (see Fig.\ \ref{diag}).

Since infrared spectroscopy features both fast probing time and sensitivity to dipolar excitations, short-range and short-lived CO fluctuations are expected to affect the low-energy optical response.  Indeed, far-infrared peaks at non-zero frequency (FIP) do appear in the real part of the in-plane conductivity $\sigma_1^{ab} (\omega)$ of LNSCO\cite{Dumm02} with $x$=0.125 below $T_{d2}$, showing their relationship with long-range charge ordering. On the other hand, similar peaks are reported also in other cuprate crystals at low temperature, independently of the technique of growth or measurement.\cite{Lupi2000,Lucarelli,Bernhard,Singley}  It is then reasonable to associate the latter FIP's with sorts of charge ordering that may not be detected by diffraction techniques due to their short range, or short life, or both.  In this respect, most observations of far-infrared anomalies concern LSCO, where such peaks were observed by several authors at energies below $\sim$ 100 cm$^{-1}$, although there is no general agreement about their domain of existence in terms of doping and temperature \cite{Lucarelli,Dumm03,Takenaka}. This situation can be partly ascribed to the difficulties in measuring a reflectivity $R(\omega) \sim 1$ at very low energy, where conventional sources provide poor brilliance, and also to the need to extrapolate $R(\omega)$ to $\omega$ = 0 for calculating $\sigma_1(\omega)$ through the Kramers-Kronig (KK) procedure \cite{josab}. 
Further information may be provided by a determination of the  $c$-axis infrared response. Above $T_c$, along the $c$ axis most superconducting cuprates behave as insulators, with $R_c(\omega) = 0.4 \div 0.7$ in the sub-THz range. Below $T_c$, Josephson tunneling of the Cooper pairs takes place between the Cu-O layers, through the non-conducting cation layer, and $R_c(\omega)$  jumps to 1. Therefore, the reflectivity change at $T_c$ is much sharper than for the $ab$ plane. An analysis of this effect, called Josephson Plasma Resonance (JPR), can provide information on the homogeneity of the charge system \cite{Dordevic}. However, due to the small condensate density along the $c$ axis, $R_c(\omega) \simeq 1$ is attained at very low infrared frequencies, such as 15 cm$^{-1}$ for LSCO with $x=0.125$ or 6 cm$^{-1}$ for optimally doped BSCCO \cite{TajimaLNSCOc,Schade2}. One thus faces again the poor brilliance of conventional infrared sources below 20 cm$^{-1}$.  

In the present work we have measured both the $R_c(\omega)$ and the $R_{ab}(\omega)$ of La$_{1.875}$Ba$_{0.125-y}$Sr$_{y}$CuO$_4$ down to 5 cm$^{-1}$. This makes negligible the interval where the reflectivity is extrapolated and - if a far-infrared anomaly (FIP) related to CO does exist - it should allow one to obtain data on both its sides. To do that we have used, in the sub-THz interval from 5 to 30 cm$^{-1}$, Coherent Synchrotron Radiation (CSR) provided by a bending magnet of the storage ring BESSY. When working at a current of 10-20 mA in the so-called low-$\alpha$ mode \cite{Schade}, it provides an intensity more than two orders of magnitude higher than that of a conventional mercury lamp, over a continuous spectrum suitable for standard Fourier-transform spectroscopy (FT-IR). 
We selected two single crystals of LBSCO with the same hole doping (as confirmed by their spectral weight, see below) $x$ = 0.125, but with different Ba concentrations. The one with $y$ = 0.05 and $T_c$ = 10 K was in the LTT phase below $T_{d2}$ = 40 K and exhibited static CO below $T_{CO} = T_{d2}$, both in X-ray \cite{FujitaXray} and neutron \cite{FujitaPrl,FujitaPrb} diffraction spectra. The other one, with $y$ = 0.085 and $T_c$ = 31 K, was in the LTLO phase below $T_{d2} = 30$ K, but showed CO superlattice spots below a $T_{CO} \sim$ 50 K $> T_{d2}$. Furthermore, in the $y = 0.085$ sample the CO signal was detected in the X-ray pattern only, not in the neutron pattern. Since in the X-ray experiment inelastic scattering can also occur, this fact was interpreted by the authors \cite{FujitaNew2} as indication of CO fluctuations, in the absence of static CO, consistently with the second-order transition scenario. The aim of this work is therefore to monitor the effect of the crossover between static and fluctuating CO states on the low-energy electrodynamics of a cuprate.

\section{Experiment}

Single crystal rods of LBSCO were grown by the Travelling-Solvent Floating-Zone  method and fully characterized as described in Ref. \onlinecite{FujitaPrb}. They were cut to produce both $ab$ and $ac$ facets of about 8x5 mm and 5x5 mm respectively. The absence of mosaicity and the facet orientation with respect to crystal axes were checked by simultaneous X-ray diffraction and optical reflection measurements \cite{josab}. The dc resistivity $\rho_{ab}(T)$ of the $ab$ facet, measured by a standard four-points probe, is plotted in Fig.\ \ref{resist}. The $T_c$ thus determined is consistent with that determined from the magnetic susceptibility \cite{FujitaPrb}. The upturn at low-$T$ in the $y=0.05$ sample is often observed in 214-systems displaying CO, such as LNSCO \cite{Ichikawa}. The quasi-linear increase at higher $T$'s and the absolute values of $\rho(T)$ in both samples are consistent with previous data on LSCO with $x$ = 0.12 \cite{Ando}. 

The reflectivity $R_{ab}(\omega)$ was measured at quasi-normal incidence (8$^0$) from 5 K to 295 K. In order to avoid any $c$-axis leakage in the determination of $R_{ab}(\omega)$, the electric field was $s$-polarized along the $a$ axis of the $ab$ facet \cite{josab}. It was then aligned along $c$ onto the $ac$ facet to determine $R_c(\omega)$. The interval 20 $\div$ 20000 cm$^{-1}$ was studied with conventional sources, the sub-THz range between 5 and 30 cm$^{-1}$  by CSR and by a dedicated bolometer working at 1.3 K. The $ab$-plane spectra were recorded with a beam current in the storage ring of 10-20 mA, in a regime of maximum available signal-to-noise ratio, while the $c$-axis data were obtained with a beam current of about 30 mA, in a regime where the intensity of the radiation is maximized (see below). The minimum frequency was limited in both cases to 5 cm$^{-1}$ by diffraction effects from the elements of the reflectivity setup. The reference was a gold (silver) film below (above) 12000 cm$^{-1}$ evaporated \textit{in situ} onto the sample.

\begin{figure}
{\hbox{\psfig{figure=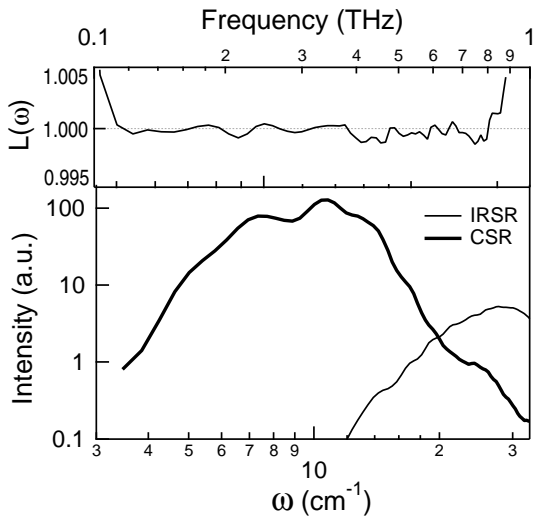,width=8cm}}}
\caption{Bottom panel: the spectrum through the normal-incidence  reflectivity  setup and the cryostat at the infrared beamline IRIS at BESSY-II in the low-alpha mode (CSR mode, stored current of 18 mA thick line) and in the normal mode (stored current of 230 mA, thin line). Top panel: the reproducibility ratio between two subsequent 256-scan spectra in the CSR mode (thin line)}
\label{ripro}
\end{figure}

\begin{figure} 
{\hbox{\psfig{figure=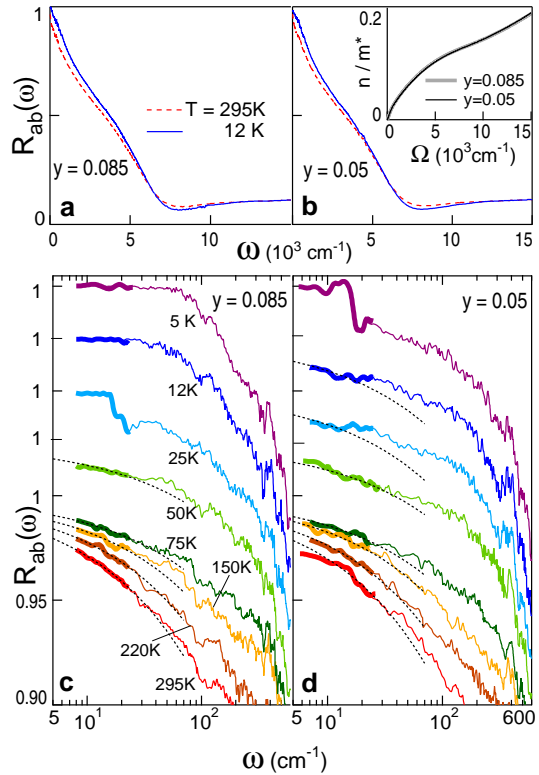,width=8cm}}}
\caption {(Color online) Reflectivity of the $ab$ plane of the La$_{1.875}$Ba$_{0.125-y}$Sr$_{y}$CuO$_4$ single crystals. Panels a,b: full range. Panels c,d: far infrared and sub-THz range. Thick lines are data taken with coherent synchrotron radiation, thin lines data collected by conventional sources. In the inset of panel b, the effective number of carriers in the conduction band is reported for both samples: the two lines can hardly be distinguished, showing that chemical doping is independent of Ba concentration.}
\label{rifab}
\end{figure}

The excellent opportunities offered by CSR as a source for Fourier-transform spectroscopy were shown and reported previously \cite{Schade2}. However, in the present work we could take advantage of several recent improvements in the CSR photon beam quality, of crucial importance for the present work. The deviations of $R_{ab}(\omega)$ from the normal metallic behavior reported in Fig.\ \ref{rifab}, which originate the FIP discussed in the present paper, are of the order of 1 \% of the $R_{ab}(\omega)$ value at the same $\omega$. It is therefore important to briefly discuss the capability of the CSR source to reach the requested level of accuracy. CSR can be affected by time-dependent effects, mainly due to instabilities in the spatial distribution of the electrons in the bunches. Time-dependent effects on the millisecond time scale may affect the baseline of the FT-IR interferograms, therefore leading to poor reproducibility in the FT-IR spectra \cite{Martin}. In the low-$\alpha$ mode developed at BESSY the electron beam instabilities have been strongly reduced by using a beam feedback system, up to stored current of about 20 mA. On the other hand, such electron beam instabilities are eventually present when the beam current is raised above 30 mA, as it was the case for the data in Fig.\ \ref{rifc}. However in the case of the $c$ axis response, the noise in the spectra does not appreciably affect the experiment. 

The reproducibility of an infrared setup is usually defined as the frequency-dependent ratio between two subsequent spectra $L(\omega)$ (also called 100 \% line). The  $L(\omega)$ of the FT-IR spectra reported in Fig.\ \ref{rifab} of the present work is shown in Fig.\ \ref{ripro} (thin line), together with one of the spectra taken on a LBSCO sample inside the cryostat (thick line). Before the experiment, a high number of subsequent interferograms were recorded by averaging over 256 acquisitions for each of them, with various scanner velocities and a spectral resolution of 1 cm$^{-1}$. Several Fourier-Transform algorithms where used to obtain the frequency spectrum, and the spectral shape was found to be robust against details of the numerical procedure. The deviations from 1 of $L(\omega)$ were found to be within 0.2 \% of the spectral intensity for $5 < \omega < 28$ cm$^{-1}$ in all cases. For completeness, we also plot the spectrum (dashed line) on the same sample and in the same experimental configuration taken with the Incoherent Infrared Synchrotron Radiation (IRSR) source, when the storage ring is run in the standard mode witha current of 150-250 mA. The spectral intensity obtained with the IRSR source is already a factor of 10 higher than that obtained with conventional sources. Therefore the gain in intensity of the CSR source over conventional sources for $\omega < 15$ cm$^{-1}$ can be estimated around 3 orders of magnitude.

\section{c-axis reflectivity}

\begin{figure} 
{\hbox{\psfig{figure=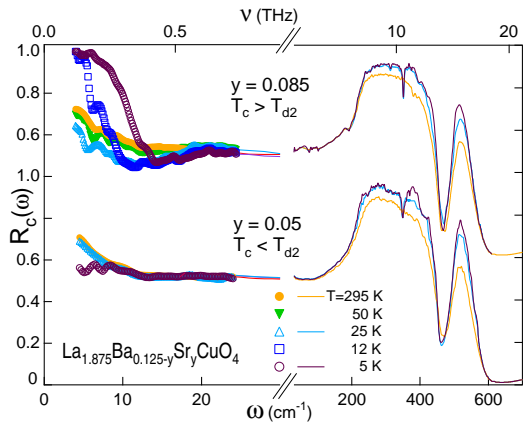,width=8cm}}}
\caption{(Color online) Reflectivity of the $c$ axis in two La$_{1.875}$Ba$_{0.125-y }$Sr$_{y}$CuO$_4$ single crystals at different temperatures. Circles refer to data taken with coherent synchrotron radiation, thin lines to data collected by conventional sources. The sample with $y$ = 0.085 exhibits, below $T_c$, a Josephson plasma resonance which displaces with temperature.}
\label{rifc}
\end{figure}

The $c$-axis reflectivity $R_c(\omega)$ is shown in Fig.\ \ref{rifc} down to 5 cm$^{-1}$. The two samples have a very similar $R_c(\omega)$ with an insulating character, namely they display phonon modes above 200 cm$^{-1}$ and, at lower frequencies, a flat $R_c(\omega) \simeq 0.6$ for any $T>T_c$. Below $T_c$ an abrupt increase of $R_c(\omega)$ in the sub-THz range is observed in the $y=0.085$ sample, but not in the $y=0.05$ sample. The reflectivity edge in the $y=0.085$ sample below $T_c$ is similar to those already reported in Ref.\ \onlinecite{Schade2} for Bi$_2$Sr$_2$Ca Cu$_2$O$_8$ (BSCCO) and in Ref.\ \onlinecite{TajimaLNSCOc} for LNSCO, and interpreted as the Josephson Plasma Resonance (JPR), indicating that the $c$ axis becomes superconducting by interlayer tunneling. 

In Fig.\ \ref{rifc} we found a JPR onset frequency $\omega_{JPR} = $ 14 cm$^{-1}$ in the $y =0.085$  sample at 5 K. The $y$ = 0.05 data at $T$ = 5 K  do not show the JPR edge  above 5 cm$^{-1}$, although they show a dip which may indicate a value of $\omega_{JPR}$ below 3 cm$^{-1}$. These values can be compared with the $\omega_{JPR} \sim$ 25 cm$^{-1}$ of Ba-free LSCO at $x$ = 0.125 \cite{TajimaLNSCOc,Dordevic}. It is apparent that the structural transition towards the LTLO or LTT phase strongly decreases the value of $\omega_{JPR}$. This observation is analogous to the shift of $\omega_{JPR}$ in LNSCO with $x=0.15$: from the Nd-free value of 60 cm$^{-1}$, $\omega_{JPR}$ decreases first to 30 cm$^{-1}$ for small Nd content, and finally below 10 cm$^{-1}$ in the LTT phase \cite{TajimaLNSCOc}. The decrease of $\omega_{JPR}$ is indicative \cite{Kleiner} of a smaller Josephson coupling between the Cu-O layers in the LTLO/LTT phase. A possible explanation, due to static CO, is a different inhomogeneous charge density in the different Cu-O layers, which reduces the overlap between the condensate wavefunctions at the origin of the Josephson effect.

The width of the JPR edge also deserves further comments. The full width $\delta_{JPR}$, measured as the difference between the onset $\omega_{JPR}$ and the frequency where $R=1$, can be estimated at $T =$ 5 K in the $y = 0.085$ sample as $\delta_{JPR}$ = 7 cm$^{-1}$. This value is the same as that of the LSCO samples with $x$=0.125 of Ref.\ \onlinecite{Dordevic}. On the other hand, typical values of $\delta_{JPR}$ for LSCO with $x \neq$ 0.125 or for optimally doped BSCCO are $\delta_{JPR} \leq$ 3 cm$^{-1}$. Following Ref.\ \onlinecite{Dordevic}, the broader JPR edge in the samples with $x$=0.125, in the framework of the two-fluid model, can be ascribed to an inhomogeneous superfluid density in the $ab$ planes, which generates a distribution of $\omega_{JPR}$. 

The $c$-axis response therefore shows that the inhomogeneous charge distribution, expected from neutron and X-ray scattering results, may coexist with the superconducting state, but strongly affects the interlayer Josephson tunneling by shifting $\omega_{JPR}$ to lower frequencies and broadening $\delta_{JPR}$. This is the case even in samples where, as here for $y$ = 0.085, $T_c$ is not appreciably reduced by the presence of CO.

\section{In-plane optical conductivity}

\begin{figure}
\begin{center} 
{\hbox{\psfig{figure=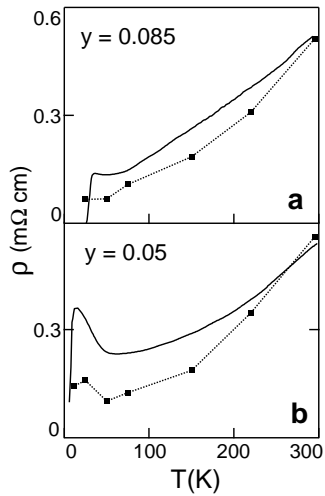,width=5cm}}}
\caption{Comparison between the in-plane dc resistivity $\rho(T)$ measured directly (solid lines) and that obtained (squares) from the Hagen-Rubens fits to $R_{ab}$ shown in Fig. 4.}
\label{resist}
\end{center}
\end{figure}

The in-plane reflectivity $R_{ab}(\omega)$ is shown at different $T$ in Fig.\ \ref{rifab} for both the $y$=0.085 and the $y$=0.05 sample, as measured with the procedure of Ref. \onlinecite{josab} and with a sensitivity of 0.2 \% in the sub-THz range (see \section{Experiment}). The two lowest optical phonons of the $ab$ plane leave a weak imprint on $R_{ab}(\omega)$ around 130 and 360 cm$^{-1}$. In the mid and near infrared $R_{ab}(\omega)$ does not show any remarkable dependence on $T$ or $y$, in the sense that it is very similar to that of Ba-free LSCO with $x \sim 0.125$  \cite{prl2005,Dumm02}. The same $R_{ab}(\omega)$ curves are reported in the bottom panels on an expanded scale, in order to appreciate how they approach 1. Therein, data taken with conventional sources (thin lines) smoothly connect with data taken with the CSR source (thick lines) at the same $T$. One observes clear signs of the SC state in the steep increase of $R_{ab}(\omega)$ to 1 around 70 cm$^{-1}$ at 5 and 12 K in the $y=0.085$ sample and around 15 cm$^{-1}$ at 5 K in the $y=0.05$ sample.

The sub-THz $R_{ab}(\omega)$ in the non-SC state was fitted by the Hagen-Rubens (HR) relation for free carriers $R_{ab} (\omega) = 1 - \sqrt{(2\omega/\pi)\sigma_{dc}^{-1}}$ which is expected to hold for $\omega<\Gamma$ where $\Gamma$ is the inverse lifetime of the carriers. In Fig.\ \ref{rifab} the obtained HR curves are reported (dashed lines) in an extended frequency range up to 70 cm$^{-1}$ at all $T$, in order to determine the energy range where they fit to data. The comparison between dashed and full lines above 20 cm$^{-1}$ in Fig.\ \ref{rifab}c-d allows one to appreciate the deviations from a standard Drude behavior in the $y=0.085$ sample for $T \leq 50$ K and in the $y=0.05$ sample for $T \leq 25$ K. These are larger than the experimental noise which is close to $\pm 10^{-3}$ (data are not smoothed). On the other hand, the same comparison also shows that a free carrier model satisfactorily describes the optical response in the LTO phase at high $T$ up to 70 cm$^{-1}$ at least. The values of $\sigma_{dc}^{-1}$ are reported in Fig.\ \ref{resist} (squares).

The deviation from a purely free carrier behavior is apparent also if one considers the in-plane dc-transport data in Fig.\ \ref{resist}. The clear upturn at low-$T$ clearly seen in the $y=0.05$ sample is expected to leave an imprint in $R_{ab}(\omega)$. However $R_{ab}(\omega)$ for any $\omega > 20$ cm$^{-1}$ monotonously increases on cooling down to 5 K. If $R_{ab}(\omega)$ would follow a  HR-like behavior at all $T$'s, then there would be a $\sigma_{dc}^{-1}$ monotonously decreasing on cooling in disagreement with the $\rho(T)$ in Fig.\ \ref{resist}-b. On the other hand, the above described fitting of the sub-THz data is in fair agreement with $\rho(T)$, but implies that the free carrier model cannot describe $R_{ab}$ for  $\omega > 20$ cm$^{-1}$. A similar argument can be used for the $y$=0.085 sample , which indeed has an almost monotonic $\rho(T)$: at 50 K, $\sigma_{dc}^{-1}$ is slightly smaller than $\rho(T)$. If the HR-fit in Fig.\ \ref{rifab}-c was to be extended above 30 cm$^{-1}$, the obtained $\sigma_{dc}^{-1}$ would be even smaller. The above deviations at low $T$ could be partially explained by the decrease of $\Gamma$ with cooling, which reduces the region of validity of the HR-fit, but this would hardly explain the kink structure seen in $R_{ab}(\omega)$ above the frequency where it deviates from the HR fit.

\begin{figure} 
{\hbox{\psfig{figure=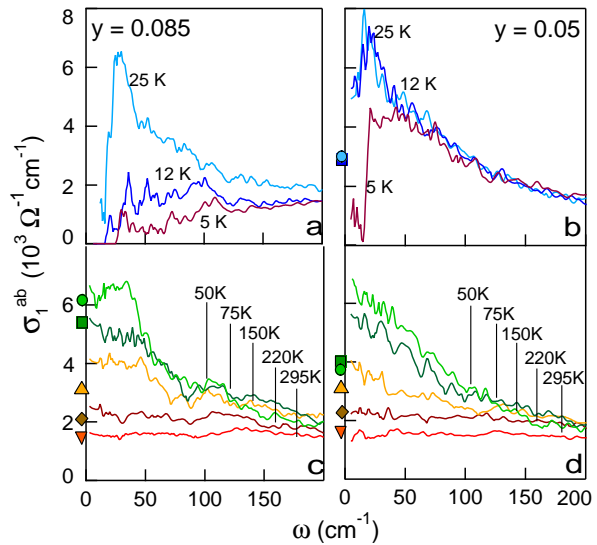,width=8.5cm}}}
\caption{(Color online) Optical conductivity of the La$_{1.875}$Ba$_{0.125-y }$Sr$_{y}$CuO$_4$ single crystals with $y$ = 0.085 (left) and  0.05 (right). Symbols in panels c,d indicate the dc conductivity values determined from $\rho(T)$ at different $T$: downward triangle = 295 K, diamond = 220 K, upward triangle = 150 K, square = 75 K, dot = 50 K. In panel b, dot = 25 K, square = 12 K.}
\label{sigma}
\end{figure}

The optical conductivity $\sigma_1^{ab} (\omega)$ was calculated by usual Kramers-Kronig transformations from the $R_{ab}(\omega)$ of Fig.\ \ref{rifab}, extrapolated between 5 and 0 cm$^{-1}$ with the above HR fit and, below $T_c$, with $R_{ab}(\omega)$ = 1. The mid-infrared conductivity comes out to be similar to that of La$_{1.88}$Sr$_{0.12}$CuO$_4$, discussed in a recent paper \cite{prl2005}, and will not be treated here. Indeed, as shown in the inset of Fig.\ \ref{sigma}-d, the spectral weight at 25 K, integrated up to any infrared frequency $\Omega$, is independent of $y$ within 1 \%. This confirms that hole doping is the same in both samples and that charge ordering phenomena affect only the electrodynamics at very low energy.
Such region is shown in Fig.\ \ref{sigma}. In both samples, a broad background conductivity at room $T$ evolves down to 75 K (panels c-d) into a Drude term monotonously decreasing with $\omega$. The optical data are in good agreement with the dc conductivity data (dots in Fig.\ \ref{sigma}), except for $y=0.05$ and $T \alt$ 75 K, where an upturn is seen in $\rho_{ab}(T)$.  At 50 K, the $y = 0.05$ sample still shows a Drude conductivity, while a broad far-infrared peak (FIP) centered at 35 cm$^{-1}$ appears in the $y=0.085$ sample, superimposed to the Drude term. Therein at 25 K (panel a) the Drude term starts transferring weight to the SC peak at $\omega$ = 0, and is well resolved from the FIP still observed at 30 cm$^{-1}$. In the $y=0.05$ sample at low $T$ (panel b), a narrow FIP shows up at 18 cm$^{-1}$ both in the 25 K and 12 K spectra. In panel b for $\omega <$ 18 cm$^{-1}$, $\sigma_1^{ab} (\omega)$ at 25 K and 12 K \emph{increases} with $\omega$, hence confirming the presence of a FIP.

In the superconducting phase at 5 and 12 K, we obtain for $y = 0.085$ $R_{ab}$ = 1.000 $\pm 0.002$ in the sub-THz range. The resulting $\sigma_1^{ab} (\omega,T\ll T_c)$, shown in Fig.\ \ref{sigma}-a, smoothly increases from zero value with increasing $\omega$ and merges with the $\sigma_1^{ab} (\omega,T=50 K)$ curve around 200 cm$^{-1}$. The absence of residual Drude absorption below $T_c$, which could appear in a $d$-wave superconductor, suggests that either it has a negligible weight, or a width smaller than 5 cm$^{-1}$, as reported for YBa$_{2}$Cu$_{3}$O$_{6+\delta}$ on the basis of GHz spectroscopy.\cite{Hardy} The absence of the FIP in the spectra of the $y=0.085$ sample at $T \ll T_c$ confirms recent observations on La$_{1.88}$Sr$_{0.12}$CuO$_4$ \cite{prl2005}. This cannot be explained by assuming that the charge distribution becomes homogeneous in the SC state, since the X-ray CO pattern remains unchanged down to a few K.\cite{FujitaNew1} One should then conclude that the opening of a gap $\Delta\sim 10$ meV along the antinodal directions of the Fermi surface (in LSCO with $x\sim 0.12$) \cite{Ino} cancels also the absorption at $\omega_{FIP} < 2\Delta/\hbar \sim 160$ cm$^{-1}$.  This clearly suggests that the FIP has an electronic, not a lattice, origin.

\section{Discussion}

The low-temperature conductivity $\sigma_1^{ab}(\omega)$ reported in Fig.\ \ref{sigma} reflects the different electronic ground state of the two LBSCO samples, in spite of a similar electrodynamic response for $T > 50$ K. In the superconducting state for $T \approx 0.5 T_c$, the conductivity is depleted in the frequency range $\omega \alt 200$ cm$^{-1}$ at $y$ = 0.085 (as in LSCO with $x$ = 0.12 \cite{prl2005}), for $\omega \alt 40$ cm$^{-1}$ at $y$ = 0.05. Even if that range does not directly measure the energy gap in the present case of $d-$wave symmetry, it is related to the magnitude of the superconducting order parameter \cite{Homes}. Therefore the data in Fig.\ \ref{sigma} are consistent with the abrupt drop in $T_c$ vs. $y$ reported in Fig.\ \ref{diag}, and  caused by the Ba-induced transition  from a high-$T_c$ SC state to a static CO state. Even more, the little change at $T_c$ displayed by the optical conductivity of the $y$ = 0.05 sample in Fig.\ \ref{sigma}-b suggests that the residual $T_c \sim$ 10 K may be due to an incomplete LTO $\to$ LTT transition, while the LTT phase is intrinsically non-superconducting. Indeed, as discussed in Ref.\ \cite{TajimaLNSCOc} for LNSCO, if sample islands remain in the LTO phase below $T_{d2}$, bulk superconductivity with a low $T_c$ may still be observed due to phase-locking among the LTO islands. 

\begin{table*}
\caption{The far-infrared peak frequency $\omega_{FIP}$ of 1/8 doped 214-systems and the temperature of the upturn in the dc resistivity,($T_{u}$), are reported in comparison with the charge-ordering ($T_{CO}$), superconducting ($T_{c}$), and structural ($T_{d2}$) transition temperatures, the in-plane lattice parameter $a$, the CO correlation length $\xi_a$, and the tilting angle of the oxygen octahedra $\theta$. Errors for $T_{CO}$ are estimated from the width of the CO transition, while for $\omega_{FIP}$ we have taken into account the uncertainty in the KK procedure.} 
\begin{ruledtabular}
\begin{tabular}{ccccccccccc}
       & y  &$\omega_{FIP}$ (cm$^{-1}$)& $T_{CO}$ (K) &$T_c$ (K) &$T_{d2}$ (K) &$T_{u}$ (K)& $a (\AA)$ & $\xi_a (\AA) $ & $\theta (^o) $\\
\hline\\
LBSCO &0.05 &14$\pm$ 2$^a$ & 37$\pm$ 1$^c$   &10$^c$  & 37$^c$    & 110$^a$   &5.355$^c$ &98$^c$  & 2.8$^c$ \\
LBSCO &0.085&35$\pm$ 5$^a$ & 50$\pm$ 5$^c$   &31$^c$  & 30$^c$    & 80$^a$   &5.327$^c$ &80$^c$  & 2.2$^c$  \\
LNSCO &0.6  &40$\pm$ 10$^b$ & 80$\pm$ 1$^c$   & 4$^c$  & 80$^c$    & 90$^b$   &5.341$^c$ &100$^c$ & 3.4$^c$  \\
\footnote{present work; $^b$Ref. \onlinecite{Dumm02},$^c$Refs. \onlinecite{FujitaPrl,FujitaPrb,FujitaXray,FujitaNew1, FujitaNew2,TranquadaPrb}.}
\label{TABLE I}
\end{tabular}
\end{ruledtabular}
\end{table*}

At temperatures not too higher than $T_c$, neither sample shows a Drude-like conductivity. As shown in the upper panels of Fig.\ \ref{sigma}, both of them exhibit a FIP similar to that previously reported for LNSCO \cite{Dumm02}. The FIP is observed in the same $T, y$ range where superlattice spots were detected by diffraction experiments: below $T_{CO} = T_{d2} = 40$ K for $y = 0.05$, around and below $T_{CO} = 50$ K for $y$ = 0.085. This is a strong indication that the FIP is intimately related to charge ordering and that it is observed for both static and fluctuating CO. The latter finding is not surprising, given the sensitivity of infrared spectroscopy to fast excitations, and confirms previous observations in Ba-free LSCO crystals \cite{prl2005}. 

At this stage, the assignment of the FIP can only be tentative and based on phenomenological considerations. Indeed, the presence of an absorption peak superimposed to the free-carrier contribution is a common feature in Charge-Density-Wave (CDW) systems with partial gapping of the Fermi surface \cite{DeGiorgi}. However, in the case of cuprates, both electron correlations and strong electron-phonon coupling are expected to complicate the conventional, BCS-like theory of CDW. Here, given the large error on the absolute value of the conductivity obtained through Kramers-Kronig transformations \cite{josab}, we choose to consider the peak frequency $\omega_{FIP}$ as the basic spectroscopic parameter. In Table I, $\omega_{FIP}$ is compared with other parameters for different 1/8 doped 214-systems, in order to look for possible correlations. Therein, $\omega_{FIP}$ varies strongly among system where both hole doping and lattice constants (\textit{e.g.}, $a$ in Table I) are the same within less than 0.5\%. Also the average octahedron tilting angle $\theta$ is clearly uncorrelated with $\omega_{FIP}$. This excludes that the different values of $\omega_{FIP}$ may be attributed to either a different density or a different environment of the ordered holes. 
 
It has been proposed \cite{Benfatto} that the FIP observed in the LSCO family is due to a collective transverse excitation of charge stripes pinned to impurities. In this framework, $\omega_{FIP}$ should reflect the size of the CO domains \cite{LeeRice}. This latter, however, is measured by the transverse CO correlation length $\xi_a$ of the diffraction experiments, which does not appreciably change among the samples of Table I. For fixed crystal parameters, charge density and domain size, the frequency of the collective mode would only be proportional to the amplitude of the charge modulation \cite{LeeRice}.  The latter quantity increases with the energy gain $\Delta U$ of the CO state with respect to a homogeneous charge distribution, which also determines the value of $T_{CO}$. Indeed, in Table I $T_{CO}$ is the only parameter which increases monotonically with $\omega_{FIP}$, hence supporting the interpretation of the FIP as a collective mode of the CO system. One may notice that, in spite of a higher $T_{CO}$ and a larger $\Delta U$ than for the static order of $y$ = 0.05, the fluctuating CO of $y$ = 0.085 does not affect $T_c$, displaying no apparent competition with superconductivity. On the contrary, collective charge excitations in the proximity of a critical transition may play a role in the pairing mechanism leading to High-$T_c$ superconductivity \cite{Caprara}. This would be the case of LSCO and LBSCO with $y=0.085$. On the other hand, in LBSCO with $y=0.05$ or in LNSCO, this cannot happen as in static CO the interaction mediated by collective charge excitations looses its singular behavior \cite{Grilli}. 

An alternative interpretation of the FIP, proposed in Ref.\ \onlinecite{Dumm02}, is based on the disorder created by Nd or Ba substitution. In disordered conductors, an increase of the carrier-impurity scattering above the localization threshold shifts the Drude peak from zero frequency to some finite-frequency value \cite{Dressel}. In the CO state, since the charges are confined in one-dimensional paths, the localization threshold would be much lower than in the homogeneous state. Therefore the Drude peak would be shifted to $\omega_{FIP}$ below $T_{CO}$. This scenario faces the difficulty that the charge carriers, lying on the Cu-O plane, are relatively insensitive to even large amount of Nd or Ba substitution. Indeed their response at higher $T$ and $\omega$ is independent of such substitution. A role of disorder in the charge transport, however, is clearly suggested by the low-$T$ upturn in $\rho(T)$, like that shown by the $y$ = 0.05 sample in Fig.\ \ref{resist}. The low-$T$ upturn cannot be directly connected to an insulating state below the structural transition at $T_{d2}$, since it starts at much higher $T$. The same holds for LNSCO \cite{Ichikawa}. Since, on the contrary, there is no upturn in the $\rho(T)$ of LSCO at 1/8 doping, that feature should be related to some sort of carrier localization induced by disorder, with no opening of a real CO-gap. Indeed, if one determines a characteristic upturn temperature $T_u$  following the procedure indicated in Ref.\ \cite{Ichikawa}, one finds $T_u$ = 80 and 110 K for $y$ = 0.05 and 0.085, respectively. $T_u$ was found \cite{Ichikawa} to be correlated with the average octahedron tilt $\theta$, which increases by increasing disorder in the cation layer (see Section II). We can therefore assume $T_u$ as a measure of the topological disorder in the sample. Once again, we remark that there is no apparent correlation between $\omega_{FIP}$ and $T_u$ or between $\omega_{FIP}$ and the number of impurity centers, thus making unlikely the interpretation of the FIP in terms of disorder-induced localization.

The dramatic effect of the rather slight structural change between the LTLO ($y > 0.075$) and the LTT ($y < 0.075$) phase on the transport and optical properties of LBSCO may appear surprising. However, recent theoretical studies have demonstrated that critical spin interactions may depend strongly on small, structural symmetry breaking \cite{Benfatto2}. In particular, the in-plane antiferromagnetic correlations of orthorombic LSCO have been found to be anomalous if compared to those of other cuprates with a tetragonal (HTT) structure like YBCO. Since in underdoped cuprates the spin degree of freedom plays a key role in the CO formation \cite{Benfatto}, we believe that the detailed understanding of the quantum transition from the high-$T_c$ SC to the static CO state at $y = 0.075$ requires further studies which should include the spin degrees of freedom.

\section{Conclusions}

In the present work we have used Coherent Synchrotron Radiation, in combination with conventional sources, to perform reflectivity measurements both along the $c$ axis and in the $ab$ plane of La$_{1.875}$Ba$_{0.125-y}$Sr$_{y}$CuO$_4$, down to 5 cm$^{-1}$. The clear observation below $T_c$ of a Josephson Plasma Resonance along the $c$-axis, besides providing a good check of the experimental efficiency in the sub-THz range, shows that the high-$T_c$ state of 1/8 doped 214-systems can survive in the presence of disorder, structural transitions and even charge ordering. We then addressed to the main purpose of the experiment, namely the study of the $ab$ plane optical conductivity in two crystals with different $y$. They were oriented and measured with a procedure which excludes any spurious contribution from the $c$ axis, often invoked to question the observations of anomalous extra-Drude contributions (FIP) in the far infrared. The two La$_{1.875}$Ba$_{0.125-y}$Sr$_{y}$CuO$_4$ single crystals were selected in order to investigate charge-order phenomena of different nature: static, and established by a structural transition for $y$ = 0.05, fluctuating and established by a second-order transition for $y$ = 0.085. In the $ab$-plane conductivity, measured for the first time in a cuprate down to 5 cm$^{-1}$, both of them showed a FIP below their different charge-ordering transitions at $T_{CO}$, which disappeared well below $T_c$. One thus finds that - in three compounds of the 214 family with evidence of charge ordering - the peak frequency increases with $T_{CO}$, then with the amplitude of the charge modulation. This latter, in turn, is not related in a manifest way to the onset of superconductivity. 

\section{Acknowledgments}

We wish to thank L. Benfatto, C. Castellani, C. Di Castro, and M. Grilli for many useful discussions. One of us (M. O.) is gratefully indebted to M. Colapietro and L. Maritato for hospitality and help during the X-ray diffraction experiments and the dc transport measurements, respectively. \\

\end{document}